\documentclass{aa}
\usepackage[varg]{txfonts}

\usepackage{graphicx}
\usepackage{caption}
\usepackage{subcaption}

\usepackage{natbib}
\bibpunct{(}{)}{;}{a}{}{,}
\usepackage[citecolor=blue, linkcolor=blue, urlcolor = black, colorlinks = true]{hyperref}
\usepackage[]{hyperref}

\usepackage{longtable}
\usepackage{lscape}

\usepackage{verbatim}
\usepackage{makecell}
\usepackage{multirow}
\usepackage{booktabs}

\usepackage{xspace}
\usepackage{comment}
\usepackage{textcomp}
\usepackage[separate-uncertainty=true]{siunitx}
\usepackage[dvipsnames]{xcolor}
\newcommand\orc[1]{\href{https://orcid.org/#1}{\includegraphics[width=3mm]{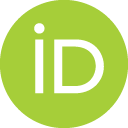}}}

\begin{document}

\title{Distributions and evolution of the  equatorial rotation velocities
  of 2937 BAF-type main-sequence stars from asteroseismology}

\subtitle{A break in the specific angular momentum at $M\simeq\!2.5\,$M$_\odot$}

\author{Conny Aerts\inst{\ref{KUL},\ref{Radboud},\ref{MPIA}}\,\orc{0000-0003-1822-7126}}

\institute{
Institute of Astronomy, KU Leuven, Celestijnenlaan 200D, B-3001
Leuven, Belgium \\
\email{conny.aerts@kuleuven.be} \label{KUL} 
\and Department of Astrophysics, IMAPP, Radboud University Nijmegen,
PO Box 9010, 6500 GL Nijmegen, The Netherlands\label{Radboud}
\and Max Planck Institute for Astronomy, K\"onigstuhl 17, 69117
Heidelberg, Germany\label{MPIA}
}

\date{Received XXX / Accepted XXX}

	\abstract
	    {Studies of the rotational velocities of intermediate-mass
              main-sequence stars are crucial for testing stellar
              evolution theory. They often rely on spectroscopic
              measurements of the projected rotation velocities,
              $V_{\rm eq}\sin\,i$. These not only
              suffer from the unknown projection factor $\sin\,i$ but
              tend to ignore additional line-profile broadening
              mechanisms aside from rotation, such as pulsations and
              turbulent motions near the stellar surface. This limits 
              the accuracy of $V_{\rm eq}$ distributions derived from
              $V_{\rm eq}\sin\,i$ measurements. }
	    {We use asteroseismic measurements to investigate the distribution
              of the equatorial rotation velocity $V_{\rm eq}$, its
              ratio with respect to the critical
              rotation velocity, $V_{\rm eq}/V_{\rm crit}$,  and 
              the specific angular momentum, $J/M$, for several thousands
              of BAF-type stars,
              covering a mass range from 1.3\,M$_\odot$ to
              8.8\,M$_\odot$ and almost the entire core-hydrogen burning phase.}
	    {We rely on high-precision model-independent internal
              rotation frequencies, as well as on masses and radii from
              asteroseismology to deduce $V_{\rm eq}$, $V_{\rm
                eq}/V_{\rm crit}$,  and $J/M$ for
              2937 gravity-mode pulsators in the Milky Way.  The sample
              stars have rotation frequencies between almost zero and
              33$\mu$Hz, corresponding to rotation periods above 0.35\,d.}
	    {We find that intermediate-mass stars experience a break
              in their $J/M$ occurring
              {in the mass interval $[2.3,2.7]\,$M$_\odot$.}
              We establish unimodal $V_{\rm eq}$ and $V_{\rm eq}/v_{\rm
                crit}$ distributions for the mass range
              {$[1.3,2.5[\,$\,M$_\odot$,} while stars with
              $M\in[2.5,8.8]$\,M$_\odot$ reveal some structure in
                  their distributions.  We find {that
                    the near-core rotation slows down as stars evolve,}
              pointing to very efficient angular momentum transport. }
	    {The kernel density estimators of the asteroseismic
              internal rotation frequency, equatorial rotation
              velocity, and specific angular momentum of this large
              sample of intermediate-mass field stars can conveniently
              be used for population synthesis studies and to
              fine-tune the theory of stellar rotation across the main
              sequence evolution.}

\keywords{Asteroseismology -- Stars: magnetic field -- Stars:
  oscillations (including pulsations) -- Stars: interiors -- Stars:
  rotation -- Stars: evolution}

\titlerunning{Asteroseismic equatorial rotation velocities and specific
  angular momenta for 2937 BAF-type stars}

\authorrunning{Conny Aerts}
\maketitle

\section{Introduction}

Single low-mass stars rotate at a pace far below their critical
velocity, $V_{\rm crit}$, defined as the velocity for which the
outwards centrifugal force at the star's equator overcomes the inward
force of gravity in a co-rotating frame of reference. In this work we
follow \citet{Aerts2019} and define low-mass stars as having a birth
mass, $M$, below 1.3 times the mass of the Sun
($M<1.3\,$M$_\odot$). All these single stars experience a so-called
Kraft break \citep{Kraft1967}.  This break is caused by angular
momentum loss due to a magnetised wind, which is fed by a dynamo in
the convective envelope.  Although high-precision space photometry
revealed the magnetic braking to stall somewhat during the second half
of the main sequence \citep{vanSaders2016,Hall2021,Metcalfe2022}, the
rotational spin-down is very effective due to the magnetic activity of
the star, particularly in the early stages after birth
\citep{Kawaler1988}.

The effective spin-down of low-mass stars caused by the magnetised
wind gives rise to the practical age-dating tool of gyrochronology,
originally introduced by \citet{Skumanich1972} and further refined by
many studies since then \citep[e.g.\,][]{Barnes2003}.  Age-dating from
gyrochronology via measurement of the surface rotation period got a
major upgrade since modern high-precision uninterrupted space
photometric light curves became available
\citep[e.g.\,][]{Barnes2010,Meibom2015}, particularly when
cross-calibrated by independent methods such as modelling of young
open clusters \citep[e.g.\ the series of papers
  by][]{Fritzewski2020,Fritzewski2021,Fritzewski2023,Fritzewski2024a}.

In terms of the range in stellar mass, the Kraft break is {quite} sharp: it
becomes ineffective when crossing a narrow mass range of 
$\sim\!0.1\,$M$_\odot$ wide \citep{BeyerWhite2024}, separating the
slowly rotating low-mass stars from the intermediate-mass stars.
Again following \citet{Aerts2019}, we define intermediate-mass stars
as having a mass in the range $1.3\,$M$_\odot<M<8\,$M$_\odot$.  This
split up between the two groups of stars agrees well with the physical
distinction caused by the Kraft break occurring {in the mass range
  $[1.3,1.4]\,$M$_\odot$ \citep{BeyerWhite2024}.
We do point out that \citet{BeyerWhite2024} deleted young field
  stars from their sample, while these stars' higher $v\sin i$ might
  be due to them being of intermediate mass rather than of low mass
  and young age. In this sense, the Kraft break could still be
  somewhat more gradual when larger samples of field stars are being
  considered.}

Intermediate-mass stars reveal a broad range of observed rotation
velocities, from hardly any rotation up to their critical velocity.
This broad coverage occurs because magnetic braking is ineffective or
absent as these stars have only a thin convective envelope or none at
all. The disappearance of the thin convective envelope and effective
spin-down in intermediate-mass stars is not known to be as sharp in
terms of the stellar mass as the classical Kraft break. Rather, the
change in behaviour of the rotational velocities and angular momenta
were found to occur between roughly $M\simeq\!1.5\,$M$_\odot$ and
$M\simeq 2.5\,$M$_\odot$, depending on the metallicity, the internal
rotation at birth, and the angular momentum loss \citep{Kawaler1988}.

From measured projected rotation velocities, $V_{\rm eq}\sin\,i$,
for stars with spectral type between B0 and F9.5 by
\citet{Fukuda1982}, and assuming a random distribution of the
inclination angle between the rotation axis and the line-of-sight,
\citet{Kawaler1987} found the specific angular momentum, $J/M$, of
dwarfs to depend on mass as $J/M\propto M^{1.09}$ when excluding Be
and Am stars, as we will do in this work.  Since that seminal
paper, numerous observational studies with spectroscopic
$V_{\rm eq}\sin\,i$ measurements for large samples of field stars
were undertaken \citep[e.g.\,][to list just a
  few]{Royer2007,Abt2009,Huang2010}.

Spectroscopic measurements are readily available from large surveys
and are often considered to deduce the rotation periods of
intermediate-mass stars. Such stars experience diverse variable
phenomena, notably large-scale oscillations \citep[see the samples by
][the latter providing a thorough
  overview]{Balona2011,Balona2013,Balona2015,VanReeth2015b,Papics2017,
  Bowman2018,Antoci2019,GangLi2020,Szewczuk2021,Kurtz2022}. The
variability caused by the multiperiodic oscillations is often dominant
over rotational modulation caused by spots in photometric light
curves.  This makes the derivation of the surface rotation periods
from such data much harder than for low-mass stars, where the
rotational modulation caused by the magnetic activity dominates the
photometric variability.

In this study we shed new light on stellar rotation from
asteroseismic measurements of the cyclic near-core rotation frequency,
denoted here as $f_{\rm rot}$, and of the accompanying equatorial
rotation velocity, $V_{\rm eq}$, for almost 3000 field stars covering
the mass range $M\in [1.3,8.8]\,$M$_\odot$.  This includes stars
having thin convective shells in their otherwise radiative outer
envelope due to partial ionisation of helium or iron-like isotopes,
which occur at temperatures of $\simeq$40kK and $\simeq$200kK,
respectively.  These ionisation layers cause large-scale low-degree
oscillation modes excited by the opacity mechanism
\citep{Pamyatnykh1999} or by flux blocking
\citep{Guzik2000,Dupret2005}.  Turbulent pressure may also occur in
the envelope and may additionally cause small-scale high-degree
oscillations \citep{Grassitelli2015,Grassitelli2016}. This gives rise
to a rather complex envelope structure.  Due to their thin convective
envelope, intermediate-mass stars are found to be incapable of
creating or sustaining an effective magnetic dynamo as their low-mass
analogues can. Hence they keep their initial rotation velocity
throughout most of their main-sequence phase and only slow down
appreciably as of their hydrogen shell burning phase
\citep{Huang2010,ZorecRoyer2012,Aerts2019,Aerts2021}.

Our aims are to provide distributions and to study the evolution of
the internal rotation frequency and the equatorial rotation velocity
of intermediate-mass stars, without having to rely on uncertain
spectroscopic $V_{\rm eq}\sin\,i$ estimates. Rather, we use
asteroseismology of gravity-mode pulsators in the Milky Way. These
field stars occur along the main sequence in the
Hertzsprung-Russell diagram from the bottom of the instability strip of
{the
$\gamma\,$Doradus ($\gamma\,$Dor hereafter)} stars \citep[][$M>1.2\,$M$_\odot$]{Dupret2005}
all
the way into the strip of the slowly pulsating B (SPB) stars
\citep[][$M<9\,$M$_\odot$]{Szewczuk2017,Pedersen2020}. Modern space
data revealed these pulsators to cover the entire main sequence in
terms of evolutionary stage, without any interruption in mass, as
highlighted by
\citet{DeRidder2023,Aerts2023,HeyAerts2024,Mombarg2024}. They are
hence a suitable population to calibrate the theory of stellar
rotation and angular momentum for the intermediate-mass regime.

\section{The two asteroseismic samples}

\begin{figure}
  \phantom{a}\vspace{-1.5cm}
  \centering
  \rotatebox{270}{\includegraphics[width=7.75cm]{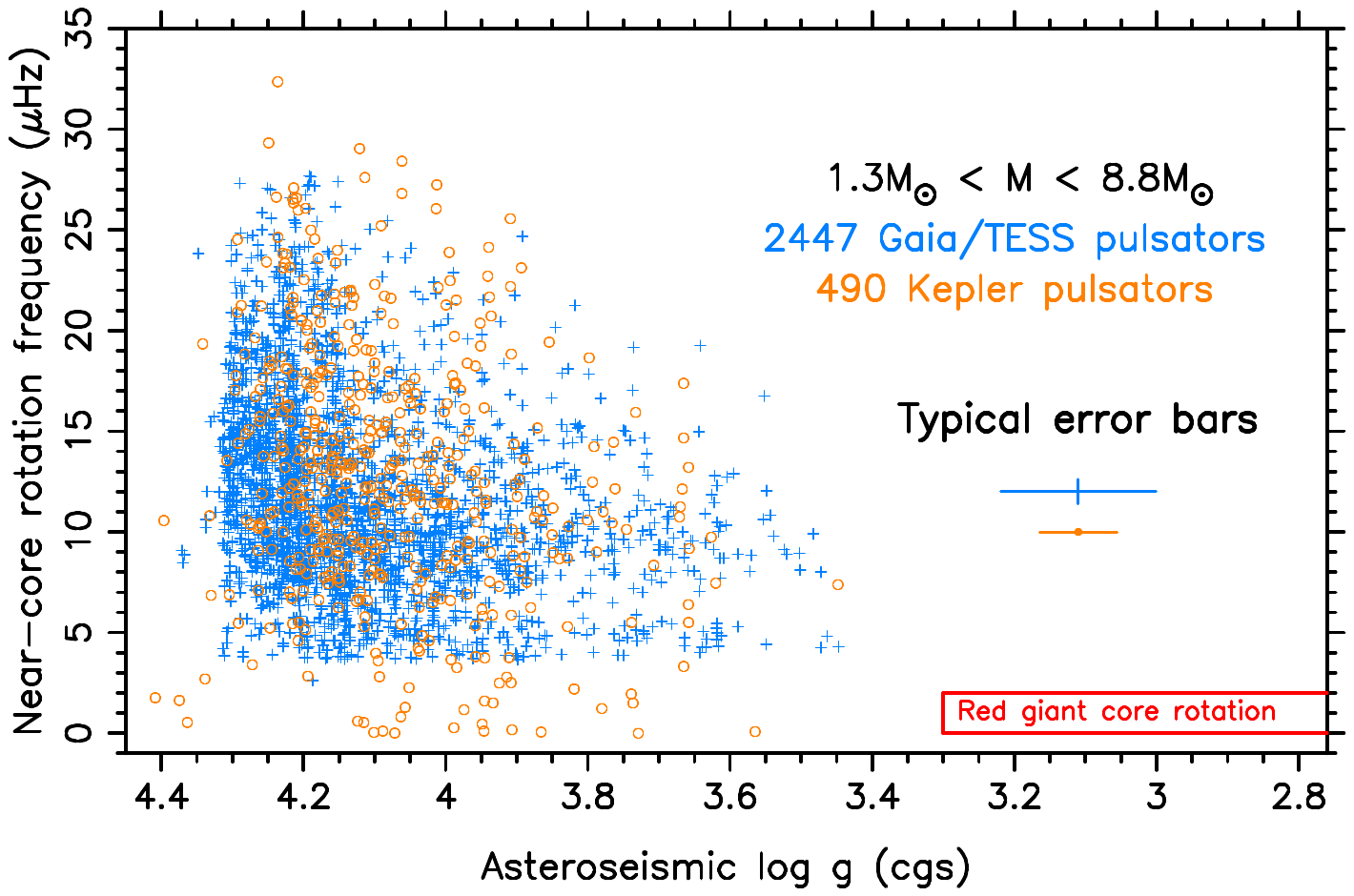}}\\[-0.7cm]
\caption{Near-core rotation frequency $f_{\rm rot}$ versus
  asteroseismic gravity for two samples: 490 {\it Kepler\/} gravity-
  and gravito-inertial mode pulsators with prograde modes from
  \citet[][orange circles]{Fritzewski2024b} and 2447 gravito-inertial
  prograde dipole mode pulsators from \citet[][blue
    crosses]{Aerts2025}. The position of the same quantities for more
  than a thousand red giant stars from mixed-mode asteroseismology is
  indicated by the red rectangle produced from
  \citet[][Fig.\,4]{Aerts2019}. }
 \label{frotlogg}
\end{figure}

We considered isolated intermediate-mass stars in the Milky Way
occurring {mostly} above the Kraft break,
{although some of them have a mass
  between 1.3\,M$_\odot$  and 1.4\,M$_\odot$ and thus  might
  be subject to classical spin down.}
Two substantial samples of
gravity-mode pulsators having their fundamental parameters determined
from asteroseismic grid modelling are available in the literature.
All the pulsators in these two samples have prograde dipole
oscillation modes identified from {\it Kepler\/} or Transiting
Exoplanet Survey Satellite (TESS) high-precision photometric light
curves. This space photometry led to a
measurement of their internal rotation frequency in the transition
layer between the convective core and radiative envelope following the
methods by \citet{VanReeth2015a,VanReeth2016} and \citet{Aerts2025}.
{This measurement of $f_{\rm rot}$ is deduced from period spacing
  patterns of identified modes and relies on the computation of the
  eigenvalues of Laplace's Tidal Equations \citep{Townsend2020}
  following the Traditional Approximation of Rotation
  \citep{Townsend2003,Mathis2013}.  This method assumes rigid rotation
  as well as modes with a large spin parameter, $s\equiv 2f_{\rm
    rot}/f_{\rm corot}$ with the denominator being the mode frequency
  in a frame of reference corotating with the star. These assumptions
  are well fulfilled for the high-order gravity modes in
  intermediate-mass pulsators
  \citep{VanReeth2016,GangLi2020,Aerts2021-GIW}, making the
  measurements of $f_{\rm rot}$ independent of stellar models.}  The
mass ($M$ expressed in M$_\odot$), radius ($R$ expressed in
R$_\odot$), and evolutionary stage (defined as the ratio of the
current hydrogen mass fraction in the fully mixed convective core to
the initial value, $X_{\rm c}/X_{\rm ini}$) {of the pulsators}
were deduced in a homogeneous way from {a grid of rotating stellar
  models calibrated by asteroseismology and computed by
  \citet{Mombarg2024}. These models have a fixed initial metallicity
  $Z=0.014$.  Further, the grid considers rotation rates between 5\%
  and 55\% of the critical rate and covers the mass range
  $[1.3,9.0]\,$M$_\odot$. It includes core boundary mixing by adopting
  an exponentially decaying convective core overshooting with its
  efficiency described by a free parameter. Rotational mixing in the
  envelope is also included in the models, but without any free
  parameters by relying on the theory by \citet{Zahn1992} and
  \citet{Chaboyer1992}.} 

While the asteroseismic modelling strategy was similar, these two
samples were selected from different criteria:
\begin{enumerate}
\item
  The first sample (Sample\,1 hereafter) consists of 490
  {$\gamma\,$Dor pulsators} modelled by
  \citet{Fritzewski2024b}.
  {We use their inferred masses, radii, and evolutionary stages
    from grid modelling relying on four} input observables. The first
  three are  the
  ten-base logarithm of the effective temperature ($\log(T_{\rm eff})$)
  and luminosity ($\log(L/L_\odot)$) from {\it Gaia\/} data release
  (DR)\,3, and the measured asteroseismic asymptotic period spacing $\Pi_0$
  \citep[see ][for a definition]{Aerts2010}. This seismic observable
  was deduced from {\it Kepler\/} space photometry by
  \citet{GangLi2020}, who relied on prograde gravity or
  gravito-inertial modes of consecutive radial order following the
  method developed by \citet{VanReeth2015a,VanReeth2016}. Aside from
  $\Pi_0$, \citet{GangLi2020} also measured the stars' near-core
  rotation frequency, $f_{\rm rot}$, from identified prograde
  modes.  This {fourth} observable comes with superb precision, often
  better than 1\%, irrespective of the used methodology
  \citep[][Fig.\,5]{Ouazzani2019}. This sample covers $f_{\rm rot}\in
        [0,33]\mu$Hz.
\item
The second sample (Sample\,2) is composed of 2464 gravito-inertial
prograde dipole mode pulsators distilled by \citet{Aerts2025} from
combined {\it Gaia\/} and TESS photometric light curves.  For these
stars, the dominant mode was used to estimate $f_{\rm rot}$ from the
recipe designed by \citet{Aerts2025}. This regression formula is only
valid for gravito-inertial modes in moderate and not too fast rotators
(in the definition by \citet{AertsTkachenko2024}, which we adopt here)
having a mode spin parameter $s>1$.  Hence, by construction, this
sample does not contain the slowest nor the fastest rotators among
intermediate-mass pulsators. In practice, it covers $f_{\rm rot}\in
[3,28]\mu$Hz.  The masses, radii, and evolutionary stages {of
  these 2464 pulsators were determined from the same stellar model
  grid used for Sample\,1. However, in this case these parameters were
  inferred from the {\it Gaia\/} DR3 $\log(T_{\rm eff})$ and
  $\log(L/L_\odot)$, and from $f_{\rm rot}$ as the three} observables,
because there is no asteroseismic measurement of $\Pi_0$ available for these
stars.
\end{enumerate}

As an update for \citet[][their Fig.\,6]{Aerts2021}, we show the 
  measured near-core rotation frequency as a function of the
model-dependent gravity deduced from the masses and radii in
Fig.\,\ref{frotlogg}. The figure also contains a comparison of these
quantities for our sample stars with those of more than a thousand red
giants from mixed dipole modes detected in {\it Kepler\/}
observations.  Figure\,\ref{frotlogg} visualises the difference in
selection criteria for our two samples, where \citet{Aerts2025}
restricted to pulsators having a dominant gravito-inertial mode with
spin value $s>1$.  {We also point to a lack of stars with a
  rotation frequency around 6.5$\mu$Hz. This is an artefact of the
  exclusion of stars having a dominant frequency of exactly once per
  day, which may be an intrinsic oscillation mode but
  could also be due to an instrumental alias frequency rather than being a
  stellar signal \citep[cf.\,][]{HeyAerts2024}.  Vetting such objects
  implies an underrepresentation of stars with $f_{\rm rot}\simeq
  6.5\mu$Hz after application of the recipe in
  \citet[][Eq.\,(4)]{Aerts2025}. This is particularly visible for
  stars towards the end of the main sequence. }

Figure\,\ref{frotlogg} illustrates
the spin-down of the near-core rotation during the main sequence and
towards the red giant phase for evolved stars with a mass
below $\simeq\!2.5\,$M$_\odot$.
Rather than using $\log\,g$ to indicate how far the stars are evolved
along the main sequence, we will work with the evolutionary stage
$X_{\rm c}/X_{\rm ini}$, which decreases from one at the zero-age main
sequence (ZAMS) to zero at the terminal-age main sequence (TAMS).
This proxy for the evolutionary stage is {a more} useful quantity for
the broad mass range covered by Samples\,1 and 2, irrespective of the
actual stellar age. Moreover, $X_{\rm c}/X_{\rm ini}$ is well
accessible by asteroseismic modelling based on low-frequency
gravity modes because these probe the region adjacent to the
convective core \citep{Pedersen2018,Pedersen2021,Mombarg2021,Michielsen2021}.

Eight stars belong to both samples. Their masses and radii are in
agreement to within 1$\sigma$ and differ less than 
  0.061\,M$_\odot$ and 0.235\,R$_\odot$, respectively.  For these
stars, we {took} the results for $f_{\rm rot}$ from
\citet{GangLi2020} because this measurement relies on tens of modes
rather than just the dominant one and is therefore more precise than
the one following the recipe from \citet{Aerts2025} -- see
Fig.\,\ref{frotlogg} for typical errors. We thus keep these stars in
Sample\,1 and delete them from Sample\,2.  Moreover, nine of the stars
in \citet{Aerts2025} turn out to have a too imprecise mass, radius, or
$f_{\rm rot}$ because their values would make them rotate faster than
the critical Keplerian rotation velocity, $V_{\rm crit}$
\citep{Aerts2025-Err}. We therefore excluded these nine stars from
Sample\,2, ending up with a total of 2447 gravito-inertial pulsators
from \citet{Aerts2025}. This gives us a total number of 2937 pulsators
to work with in the following sections. The median errors for the
asteroseismic masses and radii in Sample\,1 amount to 
  0.139\,M$_\odot$ and 0.116\,R$_\odot$, respectively. For Sample\,2
  these median errors are 0.049\,M$_\odot$ and 0.138\,R$_\odot$.
  \footnote{
  For reasons of reproducibility and follow-up studies, the used stellar
  parameters for the 2937 stars are provided in electronic form in
  Tables\,1 and 2 at
  the CDS.}


\begin{figure*}[t!]
\rotatebox{270}{\includegraphics[width=7cm]{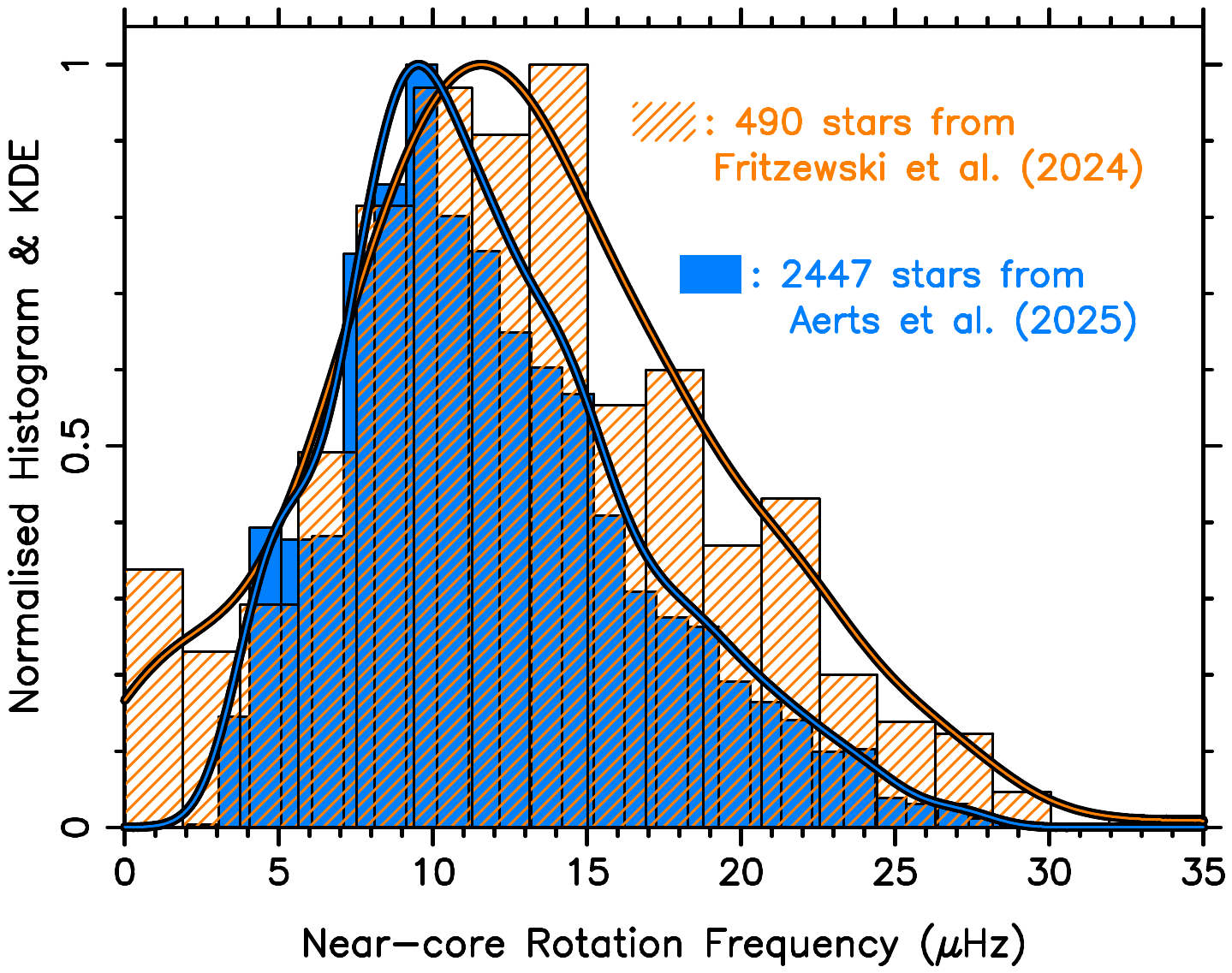}}
\rotatebox{270}{\includegraphics[width=7cm]{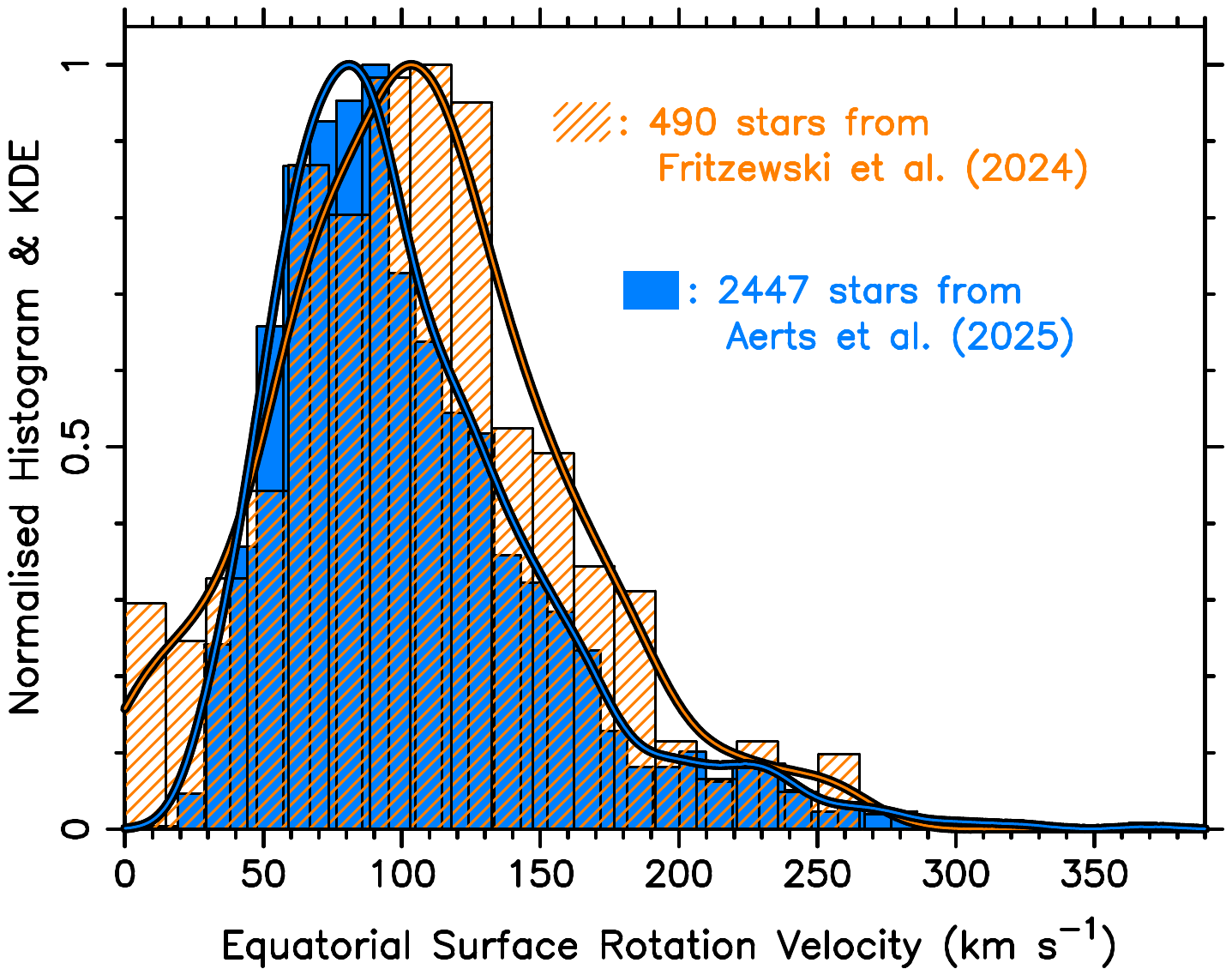}}
\rotatebox{270}{\includegraphics[width=7cm]{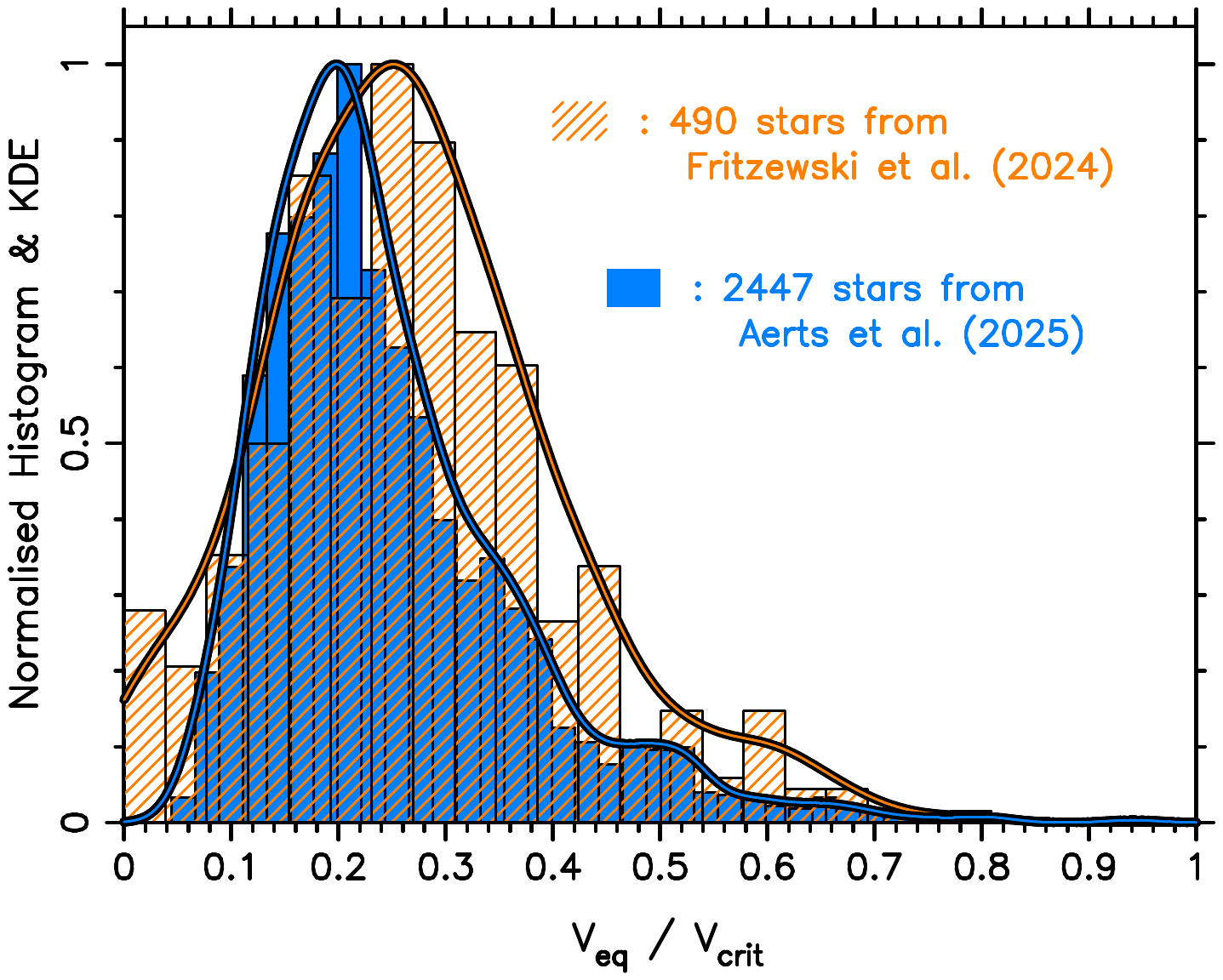}} 
\rotatebox{270}{\includegraphics[width=7cm]{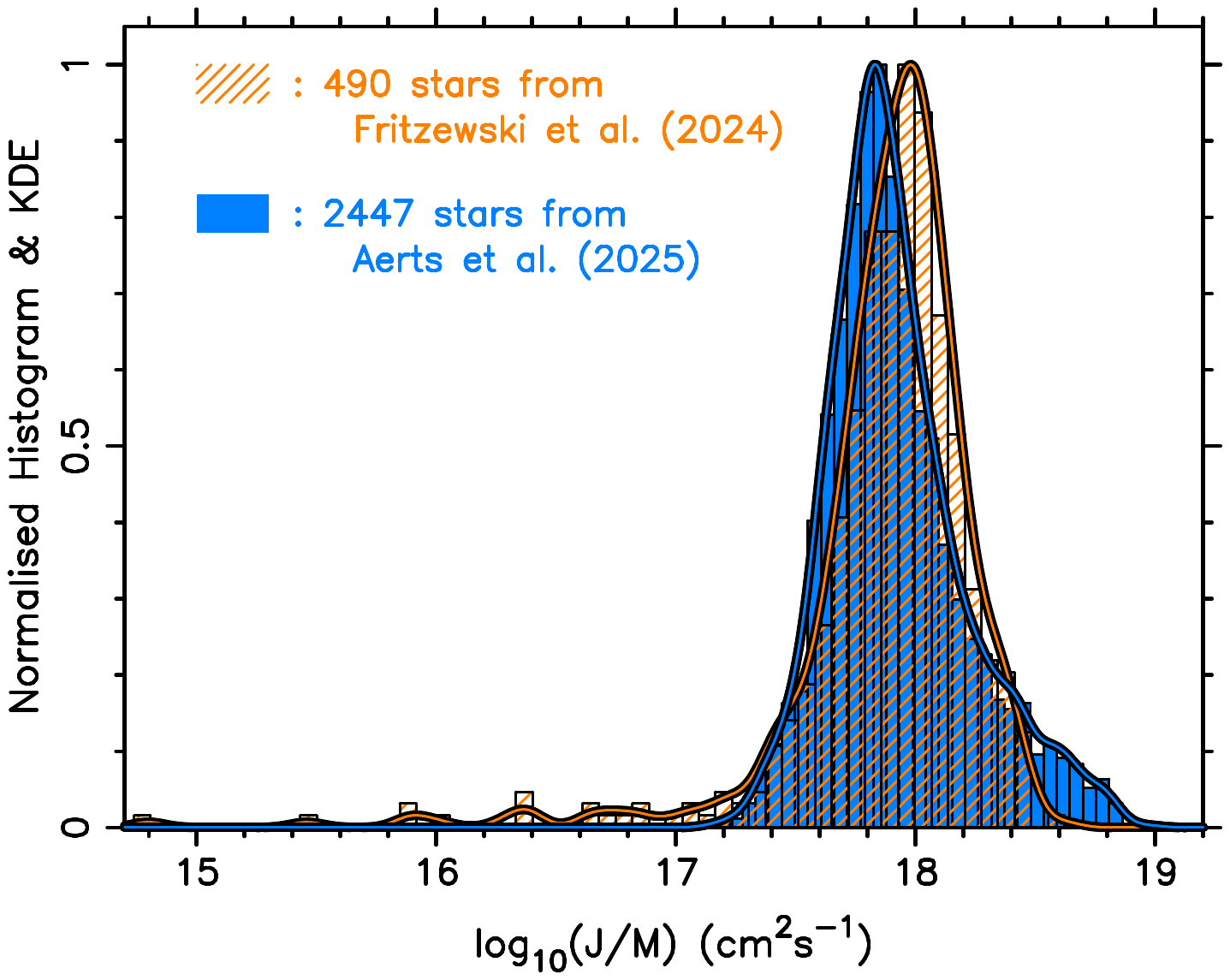}}
 \caption{Kernel density estimates (full lines) and histograms
   constructed with optimal bandwidth for Sample\,1 (orange hatched) and Sample\,2
   (blue). Upper left: model-independent measurement of $f_{\rm rot}$;
 upper right: model-dependent $V_{\rm eq}$; lower left:
 model-dependent  $V_{\rm eq}/V_{\rm crit}$; lower right: model
 dependent specific angular momentum $J/M$.}
  \label{KDEs}
\end{figure*}

\section{Equatorial rotation velocities and specific angular momenta}

Stellar rotation studies in the literature covering similarly large
samples of field stars in the galaxy as ours measured $V_{\rm
  eq}\sin\,i$ from spectral line broadening. These measurements thus
have, by construction, infinite uncertainty due to the unknown
inclination angle. This is usually corrected for by making the
reasonable assumption that the inclination angles are randomly
oriented to deduce $V_{\rm eq}$ as $2/\pi\cdot (V_{\rm eq}\sin\,i)$
{\citep{Abt2001}.}

Relying on various data sources, the landmark study by
\citet{ZorecRoyer2012} assembled $V_{\rm eq}\sin\,i$ measurements for
a sample of 2014 B6- to F2-type stars in the galaxy, covering the mass
range $[1.4,5.2]\,$M$_\odot$.  They assumed a 10\% relative error for
their assembled $V_{\rm eq}\sin\,i$ measurements, which is rather
optimistic given that many of the intermediate-mass main-sequence
stars are also subject to time-dependent spectral line broadening
mechanisms, such as oscillations, spots, or turbulent envelope
convection, to list just a few
\citep{Aerts2009,Grassitelli2015,Fremat2023,Aerts2023}.  When
these phenomena are ignored in the derivation of $V_{\rm eq}\sin i$,
the values are systematically overestimated \citep{Aerts2014}.

\citet{ZorecRoyer2012} performed their study prior to the space
asteroseismology revolution.  The latter meanwhile showed that the
theory of angular momentum transport adopted in stellar models relying
on the local conservation of angular momentum needs appreciable fixes
for stars of low and intermediate mass. This was derived from observed
dipole {gravity or mixed} modes, which offer the best probing power of the deep interiors
of stars and allow for optimal calibration of stellar evolution theory
\citep{Mosser2014,Aerts2019,Fuller2019,Eggenberger2022,Moyano2023}.

With our Samples\,1 and 2, we offer a complementary study to the one
by \citet{ZorecRoyer2012}, keeping in mind the recent asteroseismic
insights into stellar rotation and angular momentum transport.  Thanks
to the observed properties of identified dipole modes, {our} measurements
of the rotation come straight from the stellar interior, where the
evolution of stars is directed. Moreover, as the median values of the
errors for the masses and radii quoted at the end of the previous
section illustrate, these fundamental parameters from grid modelling
based on dipole-mode asteroseismology come with much smaller errors
than those resulting from evolutionary tracks in the
Hertzsprung-Russell diagram as adopted by \citet[][typically an order of
  magnitude better, see their Table\,2]{ZorecRoyer2012}.

Rather than relying on spectroscopic $V_{\rm eq}\sin\,i$ measurements,
we computed the asteroseismic equatorial rotation velocity, $V_{\rm
  eq}\equiv 2\pi\cdot f_{\rm rot}\cdot R$ for the 2937 stars in our
two samples, from the model-independent measurement of $f_{\rm rot}$
and an asteroseismically inferred radius. We did so while assuming
rigid rotation to make our study compatible with the results from
modern asteroseismology.  \citet{VanReeth2018} and \citet{GangLi2020}
have indeed shown that F-type stars are quasi-rigid rotators during
the entire main sequence, as the near-core and surface rotation
frequencies of tens of single $\gamma\,$Dor stars differ less than
10\% \citep[see Fig.\,6 in ][for a summary plot]{Aerts2021}.  More massive B-
or A-type pulsators may reveal a systematically higher level of
differential rotation between the core and envelope,
or between the envelope and surface.  Ratios of
the near-core or envelope rotation and the surface rotation have
values typically between one and two for carefully done case studies
on individual pulsators
\citep{Briquet2007,Dziembowski2008,Suarez2009,Kurtz2014,Saio2015,Triana2015,
  Salmon2022,Vanlaer2025}.  A homogeneous ensemble study with
constraints on the internal rotation from space-based asteroseismology
is available for stars at the high-mass end of our second sample. It
concerns tens of field $\beta\,$Cep stars observed with TESS and
  {\it Gaia}. This revealed almost all of them to have
envelope-to-surface rotation rates below two as well
\citep{Fritzewski2025}.  All of these asteroseismology studies imply
that quasi-rigid rotation is a valid approach for our two treated
samples of main-sequence stars. We come back to this assumption
  in Sect.\,4.

The second factor of uncertainty for $V_{\rm eq}$ is connected with
the modelled radius for our two samples. From \citet{Fritzewski2024b}
and \citet{Mombarg2024}, this has an estimated relative uncertainty of
roughly 10\% for the adopted input physics of the stellar model
grid. Overall this means that the computed asteroseismic $V_{\rm eq}$
values have similar uncertainty than the spectroscopic studies, but
are built on measurements from the stellar interior rather than the
surface \citep{Aerts2021}, and rely on rotating stellar models that
  are compatible with asteroseismic measurements.

We computed Kernel Density Estimates (KDEs) for the near-core rotation
frequency $f_{\rm rot}$, the equatorial surface velocity $V_{\rm eq}$,
the ratio of $V_{\rm eq}$ and the critical Keplerian rotation velocity
$V_{\rm crit}\equiv \sqrt{G\,M/R^3}$, and the specific angular
momentum $J/M=2/3\cdot (2\pi\cdot f_{\rm rot})\cdot R^2$ with
the {\tt  statsmodels} library of routines
\citep{Seabold2010}
available in the {\tt python} software package {\tt numpy}
\citep{Harris2020}.
These KDEs were calculated assuming a
normal distribution and by relying on Scott's rule \citep{Scott1979}
to deduce the optimal bandwidth. This rule gave equivalent results to
Silverman's approach \citep{Silverman1986} for bandwidth estimation.
Normalised histograms were subsequently constructed for each of the
four distributions, adopting the optimal bandwidths. These histograms
for the two samples are shown in Fig.\,\ref{KDEs}, along with the
KDEs.  We find the stars in Sample\,1, which consists only of early
F-type stars, to rotate somewhat faster than the much larger Sample\,2
composed of BAF-type stars.

By construction Sample\,1 offers broader distributions, because
Sample\,2 does not contain the slowest rotators nor covers the tail
towards the highest rotation frequencies that do occur in Sample\,1
(cf.\ Fig.\,\ref{frotlogg}).  We performed non-parametric two-sided
Kolmogorov-Smirnoff (KS) tests with the {\tt stats} modules of
  the software package SciPy \citep{Virtanen2020} to evaluate the
null hypothesis of dealing with identical underlying continuous
distributions for the two samples. The resulting $p-$values
  amounted to $1.9\cdot 10^{-5}$ for $f_{\rm rot}$, $1.3\cdot 10^{-5}$
  for $V_{\rm eq}$, $10^{-9}$ for $V_{\rm eq}/V_{\rm crit}$, and
  $1.5\cdot 10^{-9}$ for $J/M$. The four null hypotheses were thus
rejected with high levels of confidence ($p<0.0001$) such that, for
all four quantities, the underlying distributions from Samples\,1 and
2 indeed differ from each other on formal statistical grounds, in line
with their astrophysical selection criteria discussed in Sect.\,2.

Even if some structure occurs in the panels of Fig.\,\ref{KDEs}, we
find essentially unimodal distributions for the four quantities. This
stands in contrast to the results by \citet[][their
  Fig.\,6]{ZorecRoyer2012}, who claimed bimodal distributions from
their sample when dividing it into six sub-samples with overlaping
mass intervals.  This bimodality persisted when redistributing stars
of mass between 2.4\,M$_\odot$ and 3.85\,M$_\odot$ according to three
evolutionary stages deduced from stellar models (their
Fig.\,9). However, the bimodality disappeared and turned into a unimodal
distribution for the stars with mass between 1.6\,M$_\odot$ and
2.4\,M$_\odot$. Rather than relying on a model age, we consider
  the evolutionary stage as inferred from asteroseismic modelling and
  study the specific angular momentum properties of the samples in the
  next Section.

\section{A break in the specific angular momentum}

\begin{figure*}[t!]
  \centering
  \phantom{a}\\[-3.5cm]
\rotatebox{270}{\includegraphics[width=13cm]{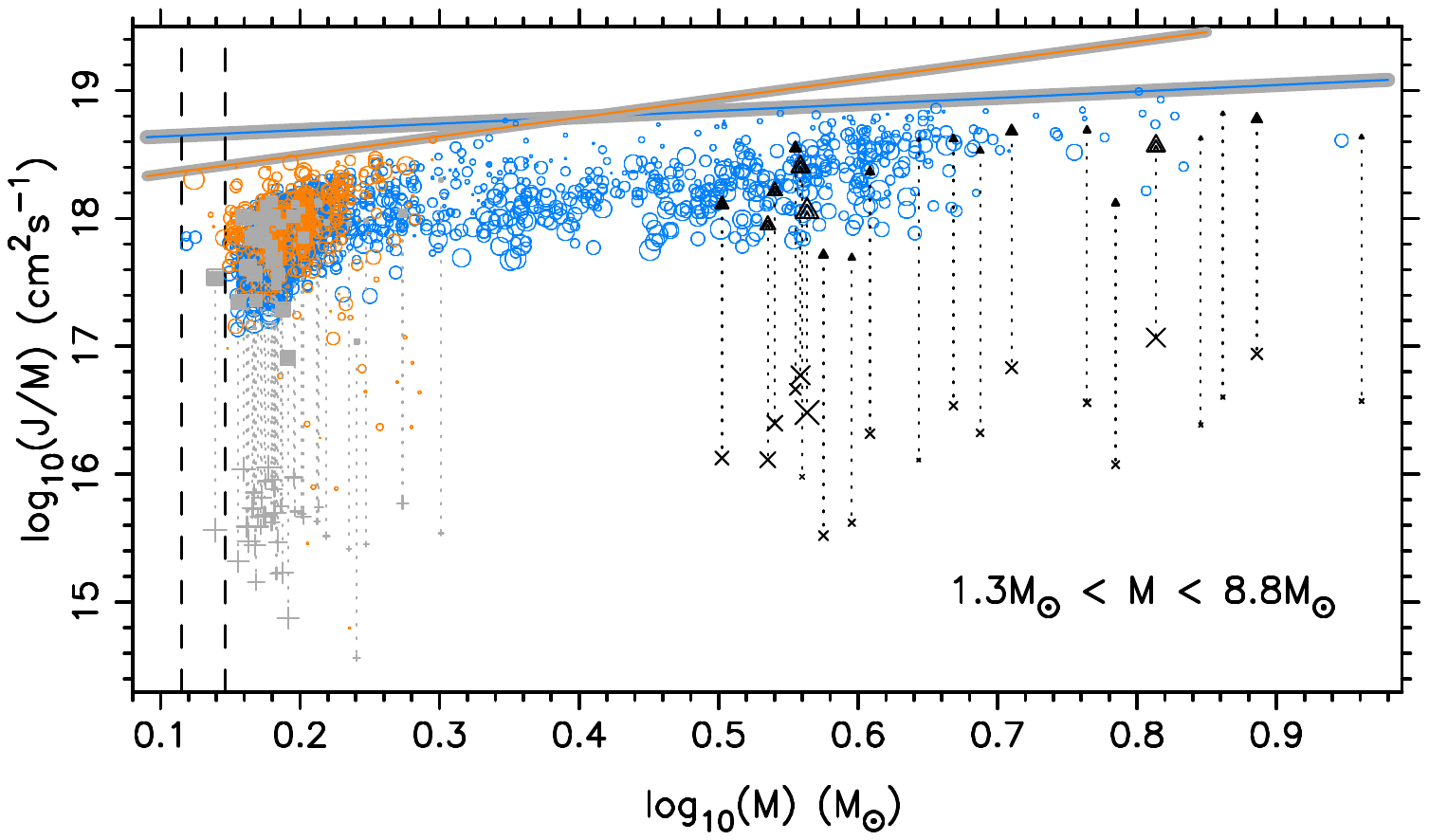}}\\[-3.5cm]
\rotatebox{270}{\includegraphics[width=13cm]{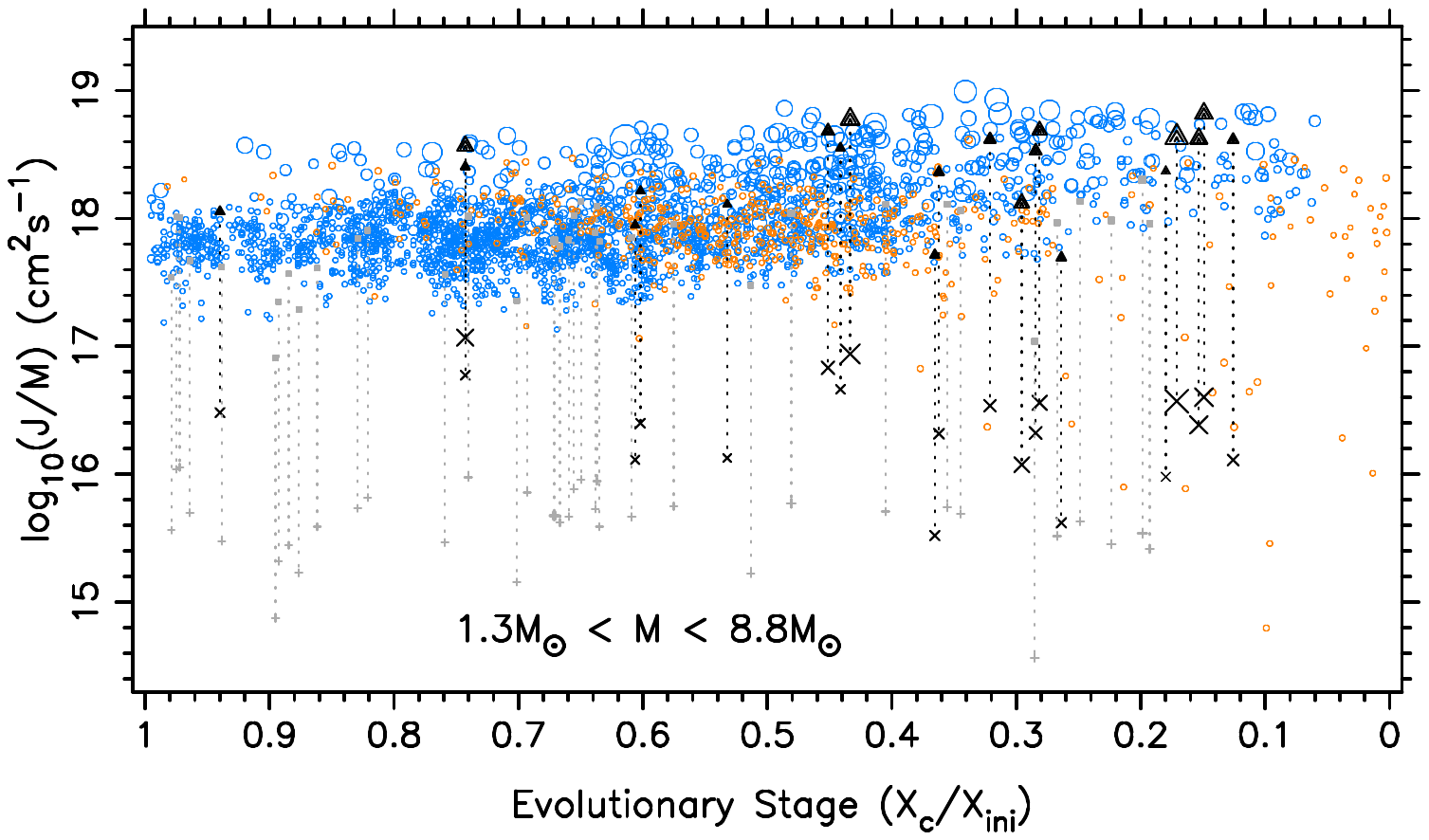}}\\[-0.5cm]
 \caption{Specific angular momentum, $J/M$, plotted logarithmically against stellar
   mass (upper panel) and evolutionary stage (lower panel). The symbol
   size for each star scales linearly with its evolutionary stage
   $X_{\rm c}/X_{\rm ini}$ (upper panel) and mass (lower panel).
   The circles show the stars of Samples\,1 and 2. Also shown
   are the $J/M$ values of 21 SPB stars from \citet[][triangles, from
     her Tables\,4 and 5]{Pedersen2022a,Pedersen2022b}, connected by a
   dotted line to the specific angular momentum of their convective
   core deduced from asteroseismic modelling of individual identified
   modes (shown as $\times$). Similarly, $J/M$ of the 37
   asteroseismically modelled $\gamma\,$Dor stars from
   \citet{Mombarg2021} are shown as filled squares connected to the
   specific angular momentum of their convective core (shown as
   $+$). {The two dashed vertical lines in the upper panel
     indicate the mass regime $[1.3,1.4]\,$M$_\odot$ where the Kraft
     break occurs.  The two coloured full lines and their uncertainty
     regions (in grey) denote the upper limits of the $J/M$
     measurements for each of the two samples.}}
\label{JM-MXc}
\end{figure*}

We now return to the concept of the Kraft break in low-mass stars,
causing their surface rotational spin-down due to the magnetic
dynamo created in the convective envelope.  It is generally
  understood that their angular momentum loss is due to a magnetised
  wind but models of the internal structure encompassing the angular
  momentum transport and loss remain uncertain \citep{Sarkar2024} and
  the transition region between early and late F-type stars is poorly
  understood \citep{Santos2025}.  As already mentioned in the
introduction, intermediate-mass stars cannot create and sustain a
large-scale magnetic dynamo as their convective envelope is either too
thin or their envelope is dominantly radiative.

Our two samples contain a few F-type stars in the transition
  region of the Kraft break but almost all of them have a mass above
  1.4\,M$_\odot$.  Figure\,\ref{frotlogg} reveals their spin-down of
the region adjacent to the convective core as the stellar evolution
progresses.  This points to the evacuation of the local angular
momentum away from the transition layer between the convective core
  and radiative envelope.  We now evaluate how $J/M$ depends on the
stellar mass and evolutionary stage, assuming that the stars keep
their rotation to be quasi-rigid, as was found for F-type pulsators by
  \citet{GangLi2020}.  

Following \citet[][Fig.\,2]{Kawaler1987} we plot the logarithm of $J/M$ versus $M$ in the upper
 panel of Fig.\,\ref{JM-MXc} for the two samples
studied here, along with those for two small additional samples of SPB
and $\gamma\,$Dor stars with the best asteroseismic modelling to
date. The forward modelling for these 58 stars is based on the
fitting of tens of individual identified gravity-mode frequencies per star,
delivering the size and mass of their convective core aside from
their global mass and radius. For these 58 stars we can therefore also
plot the spectific angular momentum of their convective core.  A first
result from Fig.\,\ref{JM-MXc} is that the total $J/M$ of these 58
pulsators is in good agreement with the one for our two large samples
whose modelling is based on global observables rather than fitting of
individual mode frequencies.  It can also be seen that $J/M$ of the
convective core is typically two to three orders of magnitude smaller
than the one of the entire star.

We further indicate the two measured upper limits for $J/M$ in
  the top panel of Fig.\,\ref{JM-MXc}, one for each sample.  The
  orange (Sample\,1) and blue (Sample\,2) straight lines result from
  the two stars with the highest measured $J/M$ value. To make these
  lines somewhat independent of the choice of one particular star, we
  indicate in grey the region covered by the upper limits when considering
  several different pairs of stars among those with the highest $J/M$
  values for each of the two samples.  We find a break in the upper limits of the
  measured $J/M$ at about 2.5$\pm$0.2\,M$_\odot$. This break occurs in
  the range $[2.3,2.7]\,$M$_\odot$ and is thus a different one than the
Kraft break in the mass range $[1.3,1.4]\,$M$_\odot$ reported by
\citet[][see the two vertical dashed lines]{BeyerWhite2024},
separating the slowly rotating low-mass stars from the entire
population of the intermediate-mass stars.  \citet{Kraft1967} 
considered 2\,M$_\odot$ as a cutoff in mass above which a different
steepening of the $J-M$ relation occurs.  Our large asteroseismic
samples reveal a break in the specific angular momentum at
  $M\simeq\!2.5\,$M$_\odot$, sub-dividing the intermediate-mass stars
into those that still have a thin convective outer envelope and those
with a pure radiative envelope aside from some thin convective shells
due to partial ionisation zones with increased opacity
\citep[][Figs\,1 and 2]{Pamyatnykh1999}.

The upper limits given by the two lines in
Fig.\,\ref{JM-MXc} represent the relationships $J\propto M^{2.48}$ for
$M\leq 2.5\,$M$_\odot$ and $J\propto M^{1.50}$ for $M> 2.5\,$M$_\odot$.
We thus recover a steepening of the specific angular momentum for
stars with a mass below the transition range
$[2.3,2.7]\,$M$_\odot$.  In comparison, \citet{Kraft1970}
found $J\propto M^{1.6}$ considering stars with spectral type earlier
than F0, picking this cutoff in spectral type rather arbitrarily. By
relying on stellar models with rigid rotation -- in retrospect a
highly modern approach proven justified by space asteroseismology --
\citet{Kawaler1987} deduced $J\propto M^{2.1}$ for stars with
spectral types from B0 to F9.5, excluding Ap and Be stars as
is the case for our two samples. We have now sharpened these old
limits and clarified the value of the cutoff mass from our much bigger
samples with asteroseismically inferred $J/M$ values.

In terms of evolutionary stage (lower panel of Fig.\,\ref{JM-MXc}), we
find that $J/M$ hardly changes or decreases for most stars 
  with a mass below the transition region of $[2.3,2.7]\,$M$_\odot$.
We note that several stars in Sample\,1 are outliers in terms of lower
$J/M$ values (see also the tail towards lower values in the lower
right panel of Fig.\,\ref{KDEs}). Most of these stars occur during the
second part of the main sequence and could be merger products or
stripped stars resulting from interacting binaries.  Further we
  notice that many of the stars in Sample\,2 with a mass above the
  transition region of $[2.3,2.7]\,$M$_\odot$ show an increasing trend
  in $J/M$. Since stars lose angular momentum unless they accrete
  matter, the lower panel of Fig.\,\ref{JM-MXc} suggests that our
  assumption of rigid rotation in a single star may not be appropriate for all the
  stars in this mass regime.  The trend in $J/M$ for this
  sub-population of Sample\,2 is in line with the measurements of
  internal rotation for early B stars.  Such pulsators so far revealed
  internal-to-surface rotation frequencies up to a factor two
  \citep{Vanlaer2025,Fritzewski2025}. Such an
  internal-to-surface rotation ratio would imply a
  shift of value 0.3 downwards in $\log_{10}(J/M)$ in Fig.\,\ref{JM-MXc}.

\begin{figure*}[t!]
\rotatebox{270}{\includegraphics[width=7cm]{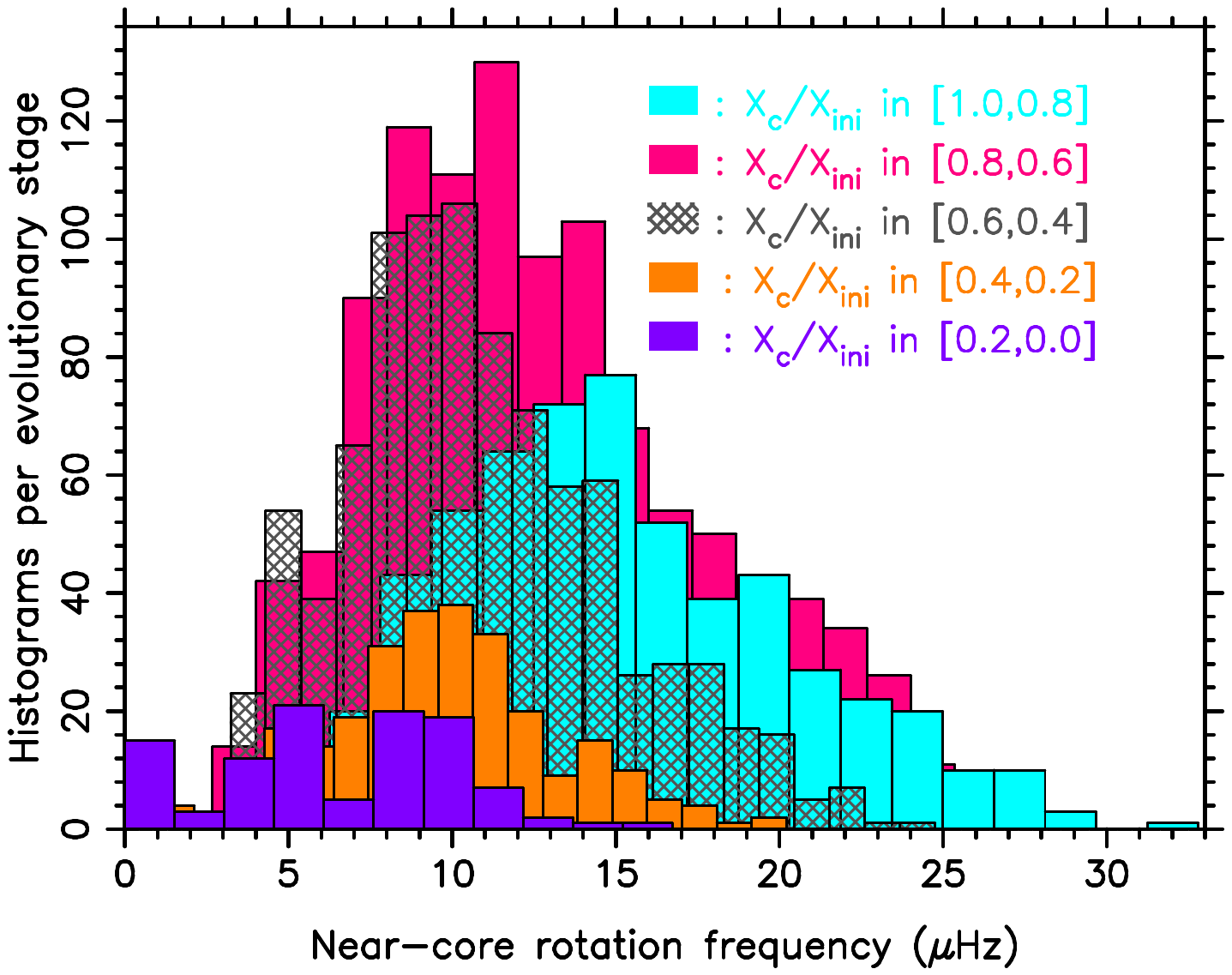}}
\rotatebox{270}{\includegraphics[width=7cm]{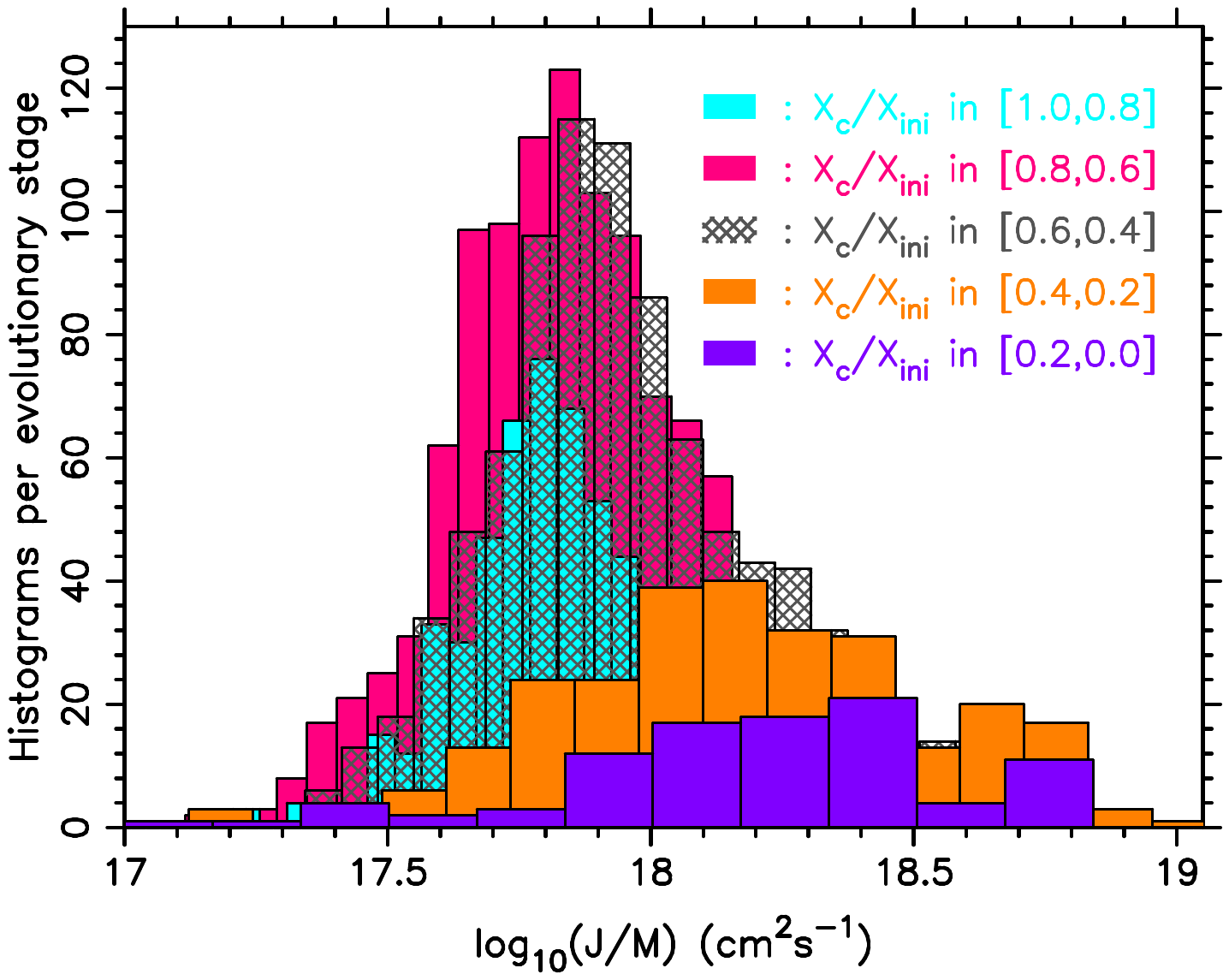}}
\rotatebox{270}{\includegraphics[width=7cm]{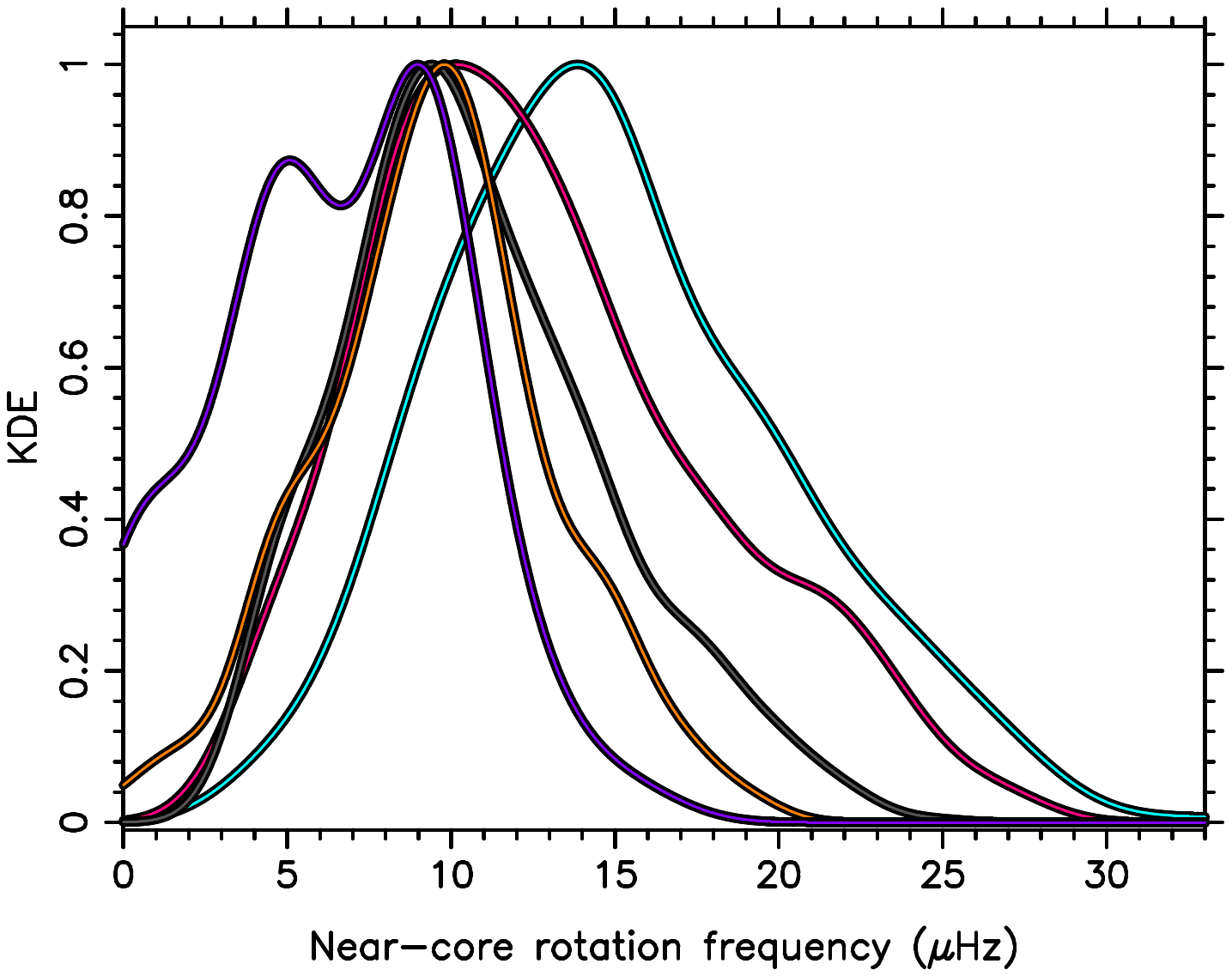}}
\rotatebox{270}{\includegraphics[width=7cm]{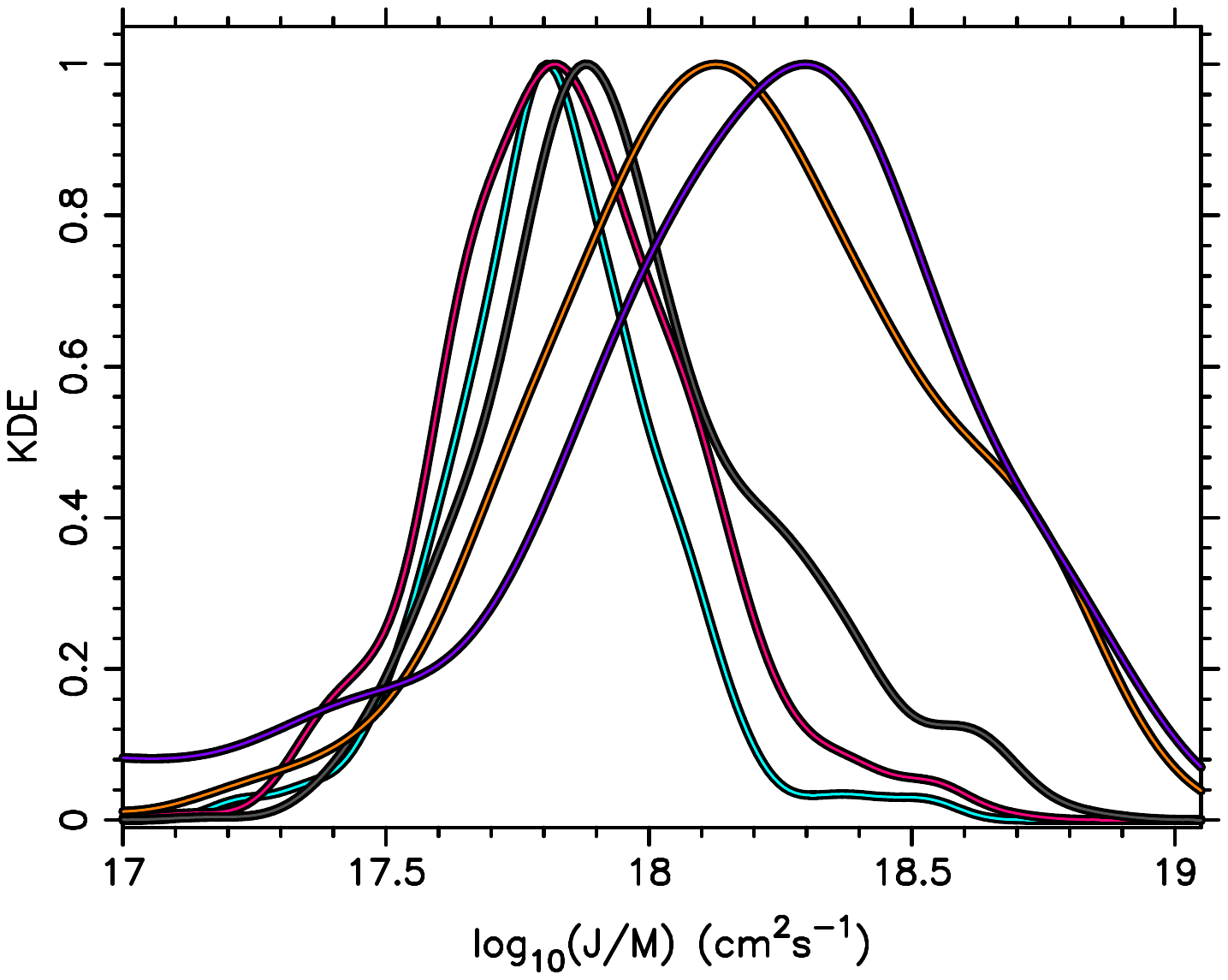}}
\caption{{Histograms (upper panels) and KDEs (lower panels)}
    of the near-core rotation frequency (left) and
  specific angular momentum (right), where the two samples are
  considered jointly, split up into five regimes of evolutionary stage
  from near-ZAMS (cyan) to near-TAMS (purple). }
\label{FigSplit}
\end{figure*}

In order to assess the evolutionary change of $J/M$ further,
we redistributed all the stars
in the two samples into five sub-samples according to their inferred
$X_{\rm c}/X_{\rm ini}$. We show the
  histograms and KDEs of the near-core rotation frequency
and the specific angular momentum for the sub-samples in these
five evolutionary stages in Fig.\,\ref{FigSplit}, where the $y-$axes
in the upper panels show the actual numbers per bin of
evolutionary stage and the lower panels the corresponding
  normalised KDEs. The left panels of this figure show that the measured
near-core rotation slows down during the main sequence evolution, as
also illustrated in Fig.\,\ref{frotlogg}. This implies efficient
angular momentum transport away from the transition layer between the
convective core and radiative envelope, as already concluded by
\citet{Aerts2019} from smaller asteroseismic samples and from
  Sample\,2 by \citet{Aerts2025}. This is
nowadays understood in terms of the joint effects of
(magneto-)hydrodynamical processes occurring during the stars'
evolution
\citep[e.g.\,][]{Pedersen2022b,Mombarg2023,Moyano2023,Moyano2024}.  The
seismic measurements of the near-core rotation of intermediate-mass
stars stand in sharp contrast with the spin up of the cores of such
stars based on local conservation of angular momentum as was still
assumed in the stellar evolution models used by
\citet{ZorecRoyer2012}.  The lower left panel shows that a
  bimodal distribution pops up for the stars that approach
  the TAMS. This sub-sample (in purple) consists of only 58 stars and
  the bimodality is caused by the artefact already discussed in the context of
  Fig.\,\ref{frotlogg}. It is due to stars with $f_{\rm rot}\simeq
  6.5\mu$Hz being under-represented in Sample\,2, particularly in the
  later stage of the main sequence.

The right panels of Fig.\,\ref{FigSplit} illustrate that the
  overestimation of $J/M$ due to our assumption of rigid rotation
  occurs mainly for $X_{\rm c}/X_{\rm ini}<0.4$. The lower right panel
  of this figure, along with the bottom panel of Fig.\,\ref{JM-MXc}
  point out that stars with a mass above the transition range
  $[2.3,2.7]\,$M$_\odot$ develop differential rotation up to a factor
  two to three (corresponding to a shift of 0.3 and 0.5 downwards in
  $\log_{10}(J/M)$, respectively) as they move towards the TAMS during the second
  half of the main sequence.
  
\begin{figure}
\rotatebox{270}{\includegraphics[width=7cm]{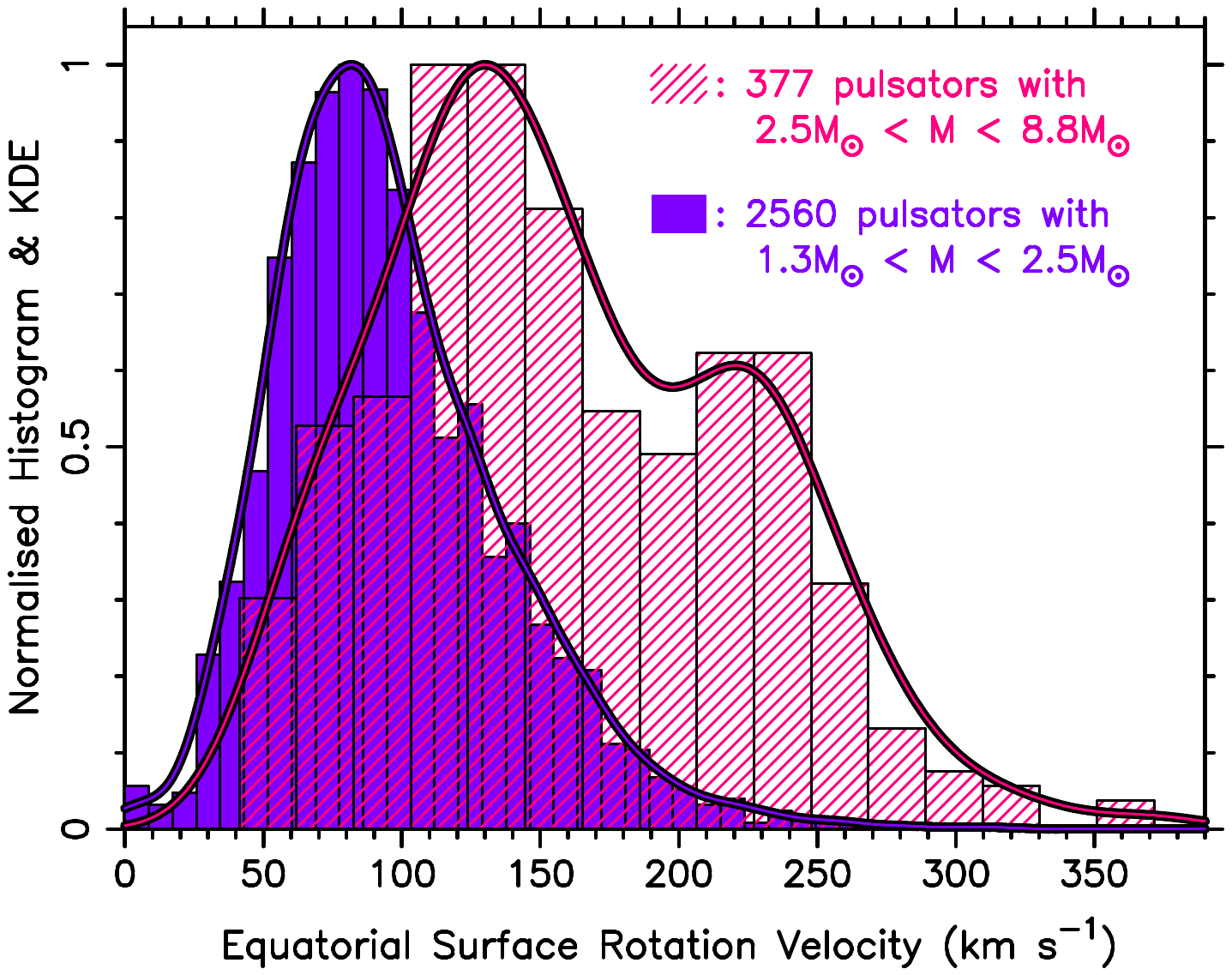}}
\rotatebox{270}{\includegraphics[width=7cm]{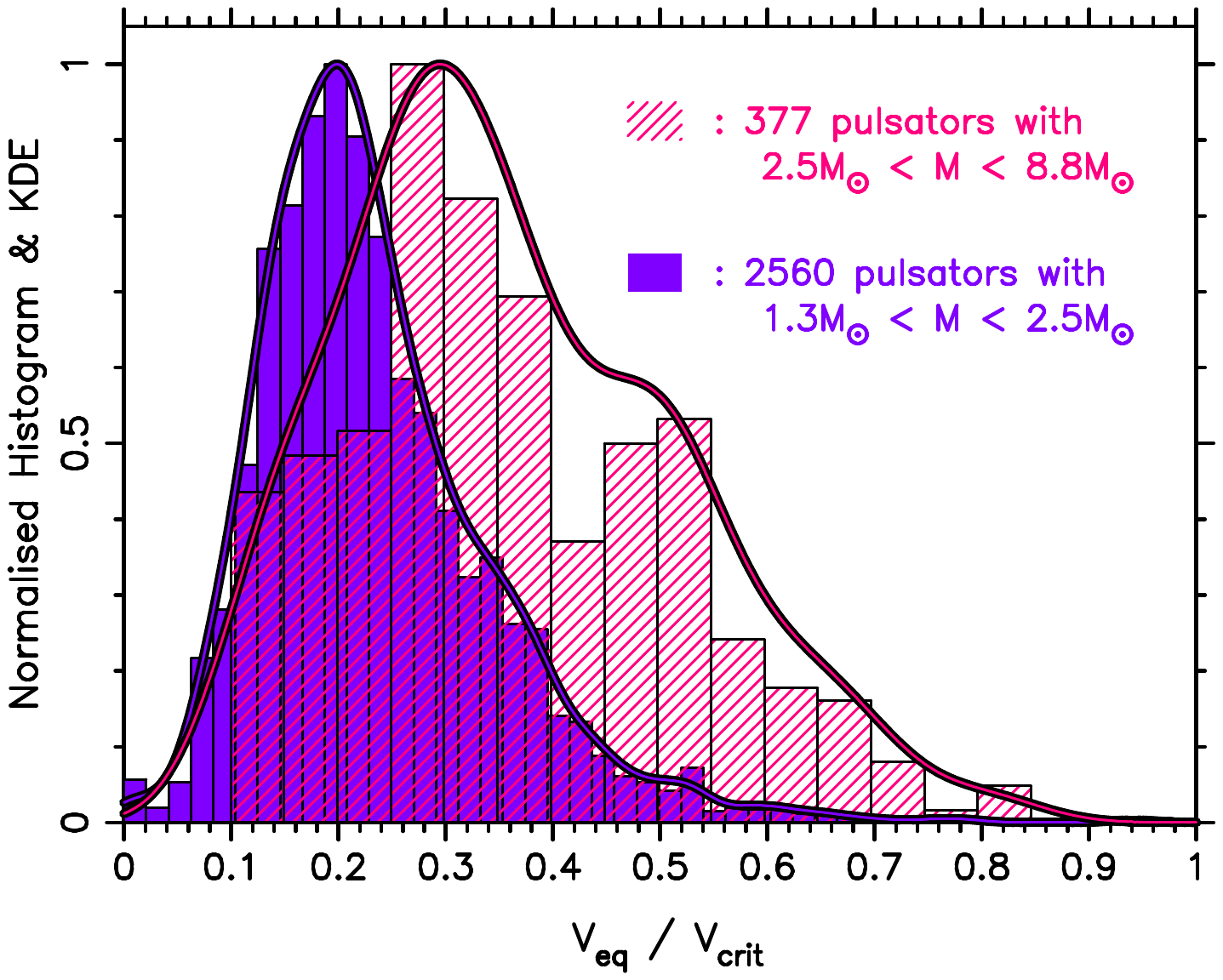}} 
 \caption{Kernel density estimates (full lines) and histograms
   constructed with optimal bandwidth for the stars
   split up according to masses below or above {2.5\,M$_\odot$.}
Upper panel: $V_{\rm eq}$; lower panel: $V_{\rm eq}/V_{\rm crit}$.}
\label{KDE-M}
\end{figure}

Given the break in $J/M$ at {$M\simeq\!2.5\,$M$_\odot$,} and the
homogeneous treatment of the two samples in terms of grid modelling,
it makes sense to re-arrange them in terms of their asteroseismic mass
and redetermine the distributions of $V_{\rm eq}$ and $V_{\rm
  eq}/V_{\rm crit}$.  We show histograms and KDEs when we group
the stars according to a mass below or above 2.5\,M$_\odot$ in
Fig.\,\ref{KDE-M}.  We observe that stars in the lower mass bin
rotate slower in absolute terms and also with respect to their
critical rate than those in the higher mass bin.  The lower and higher
mass populations peak at rotation velocities near 20\% and 30\% of the
critical velocity, respectively.

In line with \citet{ZorecRoyer2012}, who dealt with smaller sample
sizes in their spectroscopic $V_{\rm eq}\sin\,i$ study, we find
that the stars with mass above 2.5\,M$_\odot$ have a somewhat more
complex distribution of their equatorial rotation velocity compared to
the unimodal smooth distribution of the lower mass stars.  As shown in
Fig.\,\ref{JM-MXc}, the stars with $M < 2.5\,$M$_\odot$ are
characterised by a steeper $J$ versus $M$ relationship,
pointing to a stronger spin-down mechanism active in their interior.

From our two mass bins, we find hints of a dip around $V_{\rm
    eq}=170\,$km\,s$^{-1}$ along with an extra hump between $V_{\rm
    eq}=200\,$km\,s$^{-1}$ and $V_{\rm eq}=250\,$km\,s$^{-1}$,
  although this structure levels off when considering $V_{\rm
  eq}/V_{\rm crit}$.  We checked that this dip is unrelated to
  the lower density of stars with $f_{\rm rot}\simeq 6.5\mu$Hz in
  Fig.\,\ref{frotlogg} as all the stars in this $V_{\rm eq}$ range have a
  higher rotation frequency.  This small hump may simply be due to
the smaller size of the asteroseismic sample with stars having $M
  > 2.5\,$M$_\odot$, also reflected in the larger optimal bandwidth
for that sample ({20.6}\,km\,s$^{-1}$ versus 8.6\,km\,s$^{-1}$ for
the stars with {$M < 2.5\,$M$_\odot$}).  \citet{ZorecRoyer2012}
did not find such a hump at this rotational velocity but the sample
they used for the construction of the KDEs does not include stars with
$M \geq 4\,$M$_\odot$ while we have {117/377} stars in that mass
regime.

\section{Discussion and conclusions}

Thanks to the {\it Kepler\/} and TESS space missions, gravity and
gravito-inertial asteroseismology of intermediate-mass main-sequence
stars is now a mature field of research
\citep{AertsTkachenko2024}. This opens new doors for the calibration
of stellar evolution models and for population synthesis studies based
on high-precision rotational properties from asteroseismology.  We
have explored the two homogeneously analysed samples of
intermediate-mass pulsators with identified dipole gravity modes. The
distributions of their rotational equatorial velocity and specific
angular momentum were determined. Our overall sample consists of 2937
pulsators covering almost the entire main sequence. This allowed us to
unravel how the distribution of $J/M$ evolves during core-hydrogen
burning.

Guided by the asteroseismic result that the near-core rotation
frequency -- a model-independent measurement -- decreases as the stars
evolve along the main sequence, we derived a break in the $(M,J/M)$
relationship. {Stars with a mass below the transition region of
  $[2.3,2.7]\,$M$_\odot$} have a steeper relation than the more massive
stars. {This transition region} coincides roughly with the
(dis)appearance of a convective envelope for stars with sun-like
metallicity. This points towards a more effective evacuation of
angular momentum when a thin rotating convective envelope is present
compared to the case where it is absent.
In their review,
\citet{Rempel2023} conclude that small-scale dynamo fields are
expected to be active across the Hertzsprung-Russell diagram in all
stars with an outer convective zone. The authors stress that
self-consistently generated small-scale fields are only beginning to
be understood and being simulated for real circumstances.  Their
simulations for an F dwarf reveal small-scale dynamo activity, which
has an impact on the structure of the star just below the
surface. Such small-scale fields and the instabilities they cause may
be in operation in stars with $M< 2.5\,$M$_\odot$ and may be feeding a
thin magnetised wind, while being absent in stars with $M>2.5\,$M$_\odot$.
Also, from magneto-hydrodynamical simulations of rotating
  convection, \citet{Bekki2025} finds that small-scale dynamo fields have a 
non-negligible impact on angular momentum
 transport in the rotating convective envelopes of stars.

From detected gravito-inertial modes, $\gamma\,$Dor pulsators are
predicted to have stronger internal magnetic fields compared to SPB
stars \citep[][their Fig.\,8]{Aerts2021-GIW}. Considering that
3D simulations of core convection in rotating A-type stars
cause an effective magnetic dynamo in action \citep{Brun2005}, the
stronger internal fields in AF-type stars may cause more effective
angular momentum transport from the innermost regions to the envelope
layers than in B stars. These {internal and envelope} magnetic
properties acting together may imply an overall more efficient angular
momentum transport and loss for stars in the mass regime
[1.3,2.5]\,M$_\odot$ than for stars with higher masses. {We also
  found hints that stars with $M>2.5\,$M$_\odot$ deviate from rigid
  rotation as they approach the TAMS. Both phenomena may point to
  different internal magnetic properties below and above the break we
  found.}  This hypothesis on the role of large-scale internal and
small-scale envelope dynamo fields in the presence of rotation needs
to be tested further from dedicated 3D
magneto-hydrodynamical simulations covering the entire star instead of
only particular regions as is usually done for computational
convenience.  From the observational side, asteroseismic modelling
applied to gravity modes including the Lorentz force in the presence
of fast rotation as developed by \citet{Rui2024} can shed new light on
the properties of the internal magnetic fields, which were so far
  often neglected in the asteroseismic forward modelling of
intermediate-mass stars.

While our study offers alternative KDEs based on higher precision
asteroseismic measurements and inferred stellar parameters compared to
those in the spectroscopic study by \citet{ZorecRoyer2012}, it still
has room for improvements. First, the two asteroseismic samples
consist of intermediate-mass stars with sun-like metallicity while the
properties of convection depend on this parameter.  Moreover, just as
the \citet{ZorecRoyer2012} study, our two samples are limited to field
stars in the galaxy, and therefore do not cover rotation at the
stellar birth on the ZAMS and the earliest initial phases of stellar
evolution very close to the ZAMS when the stars are still in their
birth cluster. The number of stars near the TAMS is limited,
and we do not have any stars confirmed to be in the fast contraction
phase of the main sequence approaching core-hydrogen exhaustion on a
thermal time scale (stars in `the hook').  Stellar cores do
spin up in this rapid phase of evolution, as is known from the
extended main sequence turnoff stars in young open clusters
\citep[e.g.\,][]{Bastian2018}.  While spectroscopic studies of $V_{\rm
  eq}\sin\,i$ of the youngest (say, less than 300\,Myr) open clusters
are numerous \citep[e.g.\,][to list just a few recent
  ones]{Dufton2019,Kamann2023,Cristofari2025} and point to fast
rotating populations being the cause of split main sequences, our
sample of field stars does not cover the earliest evolutionary
phases and is
limited in the coverage of the fastest rotators. This is due to the
selection function of our two samples being composed of pulsators with
identified dipole gravity modes.  Asteroseismology of young open
clusters with ages below 300\,Myr recently got kickstarted
\citep{Bedding2023,Fritzewski2024a,GangLi2024,GangLi2025,Fritzewski2026} and
will be intensified in the near future with the prolonged TESS mission
and the PLATO mission on the horizon \citep{Rauer2025}. Early studies
already revealed that several clusters have fast rotators with
prograde dipole modes in their TESS data. Such asteroseismic studies
of young clusters will be of great help to extend our two samples
towards the youngest and fastest rotating intermediate-mass stars and
those close to exhausting their central hydrogen.

\section*{Data availability}
Tables\,1 and 2 with the parameters of the 2937 stars 
are available in electronic form at the CDS via anonymous ftp to
cdsarc.cds.unistra.fr (130.79.128.5) or via
http://cdsweb.u-strasbg.fr/cgi-bin/qcat?J/A+A/.

\begin{acknowledgements}
The author is grateful to Mathijs Vanrespaille, Dario Fritzewski,
Saskia Hekker, and the anonymous referee for valuable comments that
helped to improve the paper.  The research leading to these results
has received funding from the European Research Council (ERC) under
the Horizon Europe programme (Synergy Grant agreement
N$^\circ$101071505: 4D-STAR).  While partially funded by the European
Union, views and opinions expressed are however those of the author
only and do not necessarily reflect those of the European Union or the
European Research Council. Neither the European Union nor the granting
authority can be held responsible for them.
\end{acknowledgements}

\bibliographystyle{aa}
\bibliography{aa56794-25.bib}

\appendix

\end{document}